\documentclass[reprint,showpacs,preprintnumbers,amsmath,amssymb]{revtex4-1}

\pdfoutput=1
\usepackage[pdftex]{graphicx}
\pdfoutput=1
\usepackage[pdftex]{graphicx}
\usepackage[utf8]{inputenc}
\usepackage{latexsym,amsmath,amsfonts,amssymb}
\usepackage{moreverb,color,caption,multirow,footmisc}
\usepackage{hyperref}
\usepackage{bbm}
\usepackage{pbox}

\newcommand{\be}{\begin{equation}}
\newcommand{\ee}{\end{equation}}
\newcommand{\bea}{\begin{eqnarray}}
\newcommand{\eea}{\end{eqnarray}}

\newcommand{\bi}{\begin{itemize}}

\newcommand{\ei}{\end{itemize}}

\begin{document}

\title{Tighter bounds on a hypothetical graviton screening mass from the gravitational wave observation GW150914 at LIGO}
\author{Pedro Bicudo}
\email{bicudo@tecnico.ulisboa.pt}
\affiliation{Lisboa University, CFTP, Dep.\ F\'{\i}sica, Instituto Superior T\'ecnico,  Av.\ Rovisco Pais, 1049-001 Lisboa, Portugal}

\begin{abstract}
While quantum gravity is not solved yet, a screening mass for the graviton remains theoretically possible. If such a mass would screen gravity at distances of the order of the cluster galaxy radius, it could account for the universe expansion. The modified Newtonian dynamics model also could be related to a screening graviton mass at inter-galactic scales. Moreover, massive spin-2 theories constitute a very active theoretical topic. We briefly show how the very recent LIGO gravitational wave observation GW150914, emitted by a binary black hole merger distant $\sim 1.3 \times 10^9$ ly from the Earth, tightens the phenomenological bound on a hypothetical graviton screening mass, or on the effective screening of  gravity. 
\end{abstract}

\maketitle


Quantum field theory has different ways to create a mass for the interaction bosons. Thus, while the problem of quantifying gravitation remains to be solved, a screening mass for the graviton remains theoretically possible. The very recent observation the 14th of September 2015 of a gravitational wave at both the LIGO detectors at Hanford, Washington and Livingston, Louisiana,  from a binary black hole merger, GW150914
\cite{Abbott:2016blz} tightly constrains any screening of gravity. 
More detail on the LIGO observation is found in Refs. \cite{Abbott:2016jsd,Abbott:2016nhf,TheLIGOScientific:2016src,TheLIGOScientific:2016wyq,TheLIGOScientific:2016htt,TheLIGOScientific:2016uux,TheLIGOScientific:2016wfe,TheLIGOScientific:2016qqj,TheLIGOScientific:2016agk}.
Here we briefly review how the screening of gravity was phenomenologically possible before the GW150914 observation, and how the LIGO observation rules out most phenomenological interest in gravity screening or in modified Newtonian dynamics.

The discovery of the $\pi$, the lightest boson mediating the nuclear strong interaction, predicted by Yukawa \cite{Yukawa:1935xg}  in 1935, is the first experimental example of a boson screening mass, of the order of $m_{\pi^0} = 135$ MeV
\cite{Agashe:2014kda}. Nambu and Jona-Lasinio \cite{Nambu:1961tp,Nambu:1961fr} modelled chiral symmetry breaking, in 1961, leading to the mass generation of hadrons. The $\pi$ mass origin, in the Gell-Mann Oakes and Renner relation \cite{GellMann:1968rz} , is actually due to the interplay of spontaneous breaking of chiral symmetry, and to the quarks mass generation in the weak sector. More modernly, chiral symmetry breaking, leading to the $\pi$ mass relation and to other partially conserved axial current theorems, are directly related to Quantum Chromo Dynamics (QCD) with non-perturbative techniques such as the Dyson-Schwinger equations and the Bethe-Salpeter Equation \cite{Bicudo:2003fp}. 

In the electroweak sector of the Standard Model, the Higgs mechanism \cite{Higgs:1964ia,Higgs:1964pj}, not only produces masses for the fermion quarks and leptons, but also for the gauge bosons $W^\pm , \ Z$ first observed at LEP in CERN and with screening masses starting at $m_{W^\pm}=80.4$ GeV \cite{Agashe:2014kda}. The recent discovery of the Higgs boson \cite{Chatrchyan:2012xdj,Aad:2012tfa} absolutely confirmed the Standard Model of particle physics and the Higgs mechanism.

\begin{table}[t!]
\caption{\label{tab:screening} Lowest screening masses of bosons in different quantum interactions of particle physics. For the photon we assume it is massless as in the standard model, but a massive photon is experimentally possible \cite{Goldhaber:2008xy}.}
\begin{ruledtabular}
\begin{tabular}{cccc}
interaction & boson & screening mass & potential \\
\hline
strong confinement & $ g$ & 0.5 to 0.8 GeV & linear \\
weak & W, \ Z & 80 GeV & Yukawa\\
strong nuclear & $\pi , \rho $ & 0.14  GeV & Yukawa \\
electromagnetic & $\gamma$ & $ 0 $ & Coulomb
\end{tabular}
\end{ruledtabular}
\end{table}

Back to the strong interactions, at smaller scales than the hadronic interactions mediated by the $\pi$ boson, QCD also has another interesting screening mass, of a second type. This second, more microscopic, screening mass is present in the flux tubes leading to the confinement of quarks and gluons in colour singlets.   In Table \ref{tab:screening}, we separate the two different QCD scales of screening, the effective one of nuclear strong interactions related to the $\pi$ mass,  and the more fundamental one of confinement. 
Unlike the mass generation in chiral symmetry breaking and in the Higgs mechanism, the mechanism of confinement is not yet theoretically understood in full detail, and remains a difficult open problem in theoretical particle physics.
Notice the quantization of QCD, either in pure gauge QCD or in QCD with massless quarks which have no dimensional scale in the Lagrangian,  breaks conformal invariance and this was computed since the onset of lattice QCD \cite{Creutz:1979dw,Creutz:1980zw}.
Nevertheless the phenomenology of confinement is well known in lattice QCD computations of flux tubes and of the gluon propagator. The QCD flux tubes are gauge invariant and screening is related to the width of the flux tubes \cite{Cardoso:2013lla} (after accounting for the flux tube quantum vibrations) with a penetration length $\lambda \sim 0.22$ to 0.24 fm. The inverse of the penetration length
\be
\label{eq:screeningmass}
m \simeq \lambda^{-1} \ ,
\ee
indicates an effective screening mass for the gluon (or another related quantum degree of freedom) of $m \sim 0.8$ to 0.9 GeV. Besides, the transverse gluon propagator computed in Landau gauge saturates in the infrared, qualitatively consistent with a gluon screening mass of the order of $m_g \sim 0.5$ GeV in the Landau gauge \cite{Oliveira:2010xc}. Although the gluon propagator is gauge dependent, this saturation in the Landau gauge is of the same order of the saturation recently computed in different $R_\xi$ gauges \cite{Bicudo:2015rma}. The screening in QCD is most interesting because it preserves Gauss law,  and nevertheless it enhances the interaction between charges by squeezing the colour fields in flux tubes; similar to the screening of the electromagnetic field in type II superconductors \cite{Cardoso:2006mf}. 
This results in a linear potential between static charges, much stronger than the Coulomb potential. Moreover confinement screening is absolute, in the sense the gluons are unable to propagate in the vacuum. Thus, in contradistinction with the weak and nuclear types of screenings, in confinement screening waves do not propagate.

Nevertheless the quantization of a theory does not, of course, imply screening. The main case without screening is the electromagnetic sector of the Standard Model: it remains totally unscreened upon quantization. But even the photon may have a mass, See Ref. \cite{Goldhaber:2008xy} for a review of direct and indirect bounds on the photon mass. 
Assuming a propagation with no dispersion or damping in a distance with the radius of all the visible universe, with a diameter of $\sim 46 \times 10^9$ ly,  of the Cosmic Microwave Background, since the universe became transparent,  would correspond to a nearly vanishing mass  $ \ll 5 \times 10^{-34}$ eV.  The different screening masses of the quantum interactions of particle physics are listed in Table \ref{tab:screening}.

Possibly inspired in particle physics, or in condensed matter physics, the idea of a graviton mass has been considered both theoretically and in phenomenology, since the early work of Fierz and Pauli in 1939 \cite{Fierz:1939ix}. 

Theoretically, the graviton mass is a very active topic, because quantized gravity is a spin 2 theory. Inasmuch as the Proca \cite{Proca:1988ii} and Stueckelberg \cite{Stueckelberg:1900zz} theories are unavoidable to fully understand spin 1 theories, studying a graviton mass is unavoidable in the mapping of all possible sub-classes of spin 2 theories. Moreover, a tantalizing motivation to screen gravity is the gravitational constant problem \cite{Weinberg:1988cp,Jackiw:2005yc,Polyakov:2006bz,Porto:2009xj}.
The massive Fierz-Pauli theory suffers from a discontinuity
\cite{vanDam:1970vg,Zakharov:1970cc,Iwasaki:1971uz}: in the limit of a vanishing graviton mass, the Fierz–Pauli theory is not equivalent to the linearized general relativity. The discontinuity cure led to several developments of spin 2 theories, starting with the non-linear Vainshtein mechanism \cite{Vainshtein:1972sx}.

Clearly, spin 2 theories are more complex than Abelian spin 1 theories, and the massive has only been understood recently. 
Numerically \cite{Damour:2002gp}, the Vainshtein mechanism was worked out in Refs \cite{Babichev:2009jt,Babichev:2010jd}. see Ref. \cite{Babichev:2013usa} for a recent review of the Vainshtein mechanism.
Theoretically, inasmuch as a possible realization of massive spin 1 theories include a longitudinally polarized photon, in spin 2 massive theories, helicities 0 and 1 are included together with the spin 2 of general gravity. For recent theoretical reviews see Refs. \cite{Hinterbichler:2011tt,deRham:2014zqa}.
A ghost was believed to be present in massive gravity, until the model of Dvali, Gabadadze and Porrat (DGP) was proposed,  as the first theory without a ghost \cite{Dvali:2000rv,Dvali:2000hr}, and where one could understand explicitly how the Vainshtein mechanism is implemented \cite{Deffayet:2001uk,Dvali:2002vf}. In the decoupling limit of DGP 
the interactions of the helicity 0 mode become important for a finite mass, whereas the corrections to GR tend to 0 in the limit where the graviton mass goes to zero \cite{Luty:2003vm,Nicolis:2004qq}. 
The helicity 0 mode was also considered as a scalar field theory in own right, the Galileon \cite{Nicolis:2008in,Brax:2011sv}, reminiscent of the Abelian Higgs model used in the Stueckelberg photons. 
Finally, the DGP model has been successfully extended to the massive gravity theory \cite{deRham:2010ik,deRham:2010kj,deRham:2010tw} that implements the Vainshtein mechanism, fully solves the ghost problem and where the graviton is a real massive particle with a pole in the propagator.
Moreover massive gravity has several interesting properties, leading to an intense research \cite{deRham:2014zqa} 
and also inspiring solutions to problems of other approaches, such as the di-metric theories \cite{Hossenfelder:2008bg,Hassan:2011hr}.

We now discuss the phenomenological motivation to a hypothetical graviton screening mass
\cite{Goldhaber:1974wg}, which is appealing at two different scales. Notice at distances of the solar systems and smaller, Newtonian gravity is correct in first order, and general gravity is 100\% compatible with experiment in all observations, and thus is assumed to be correct. Newtonian dynamics is a first order approximation to general gravity, and is a direct result of the Gauss law. Thus no screening exists at this scale, and screening is only phenomenologically interesting at much larger scales.

\begin{figure}[t!!]
\includegraphics[trim=10 10 10 10,clip,width=0.9\columnwidth]{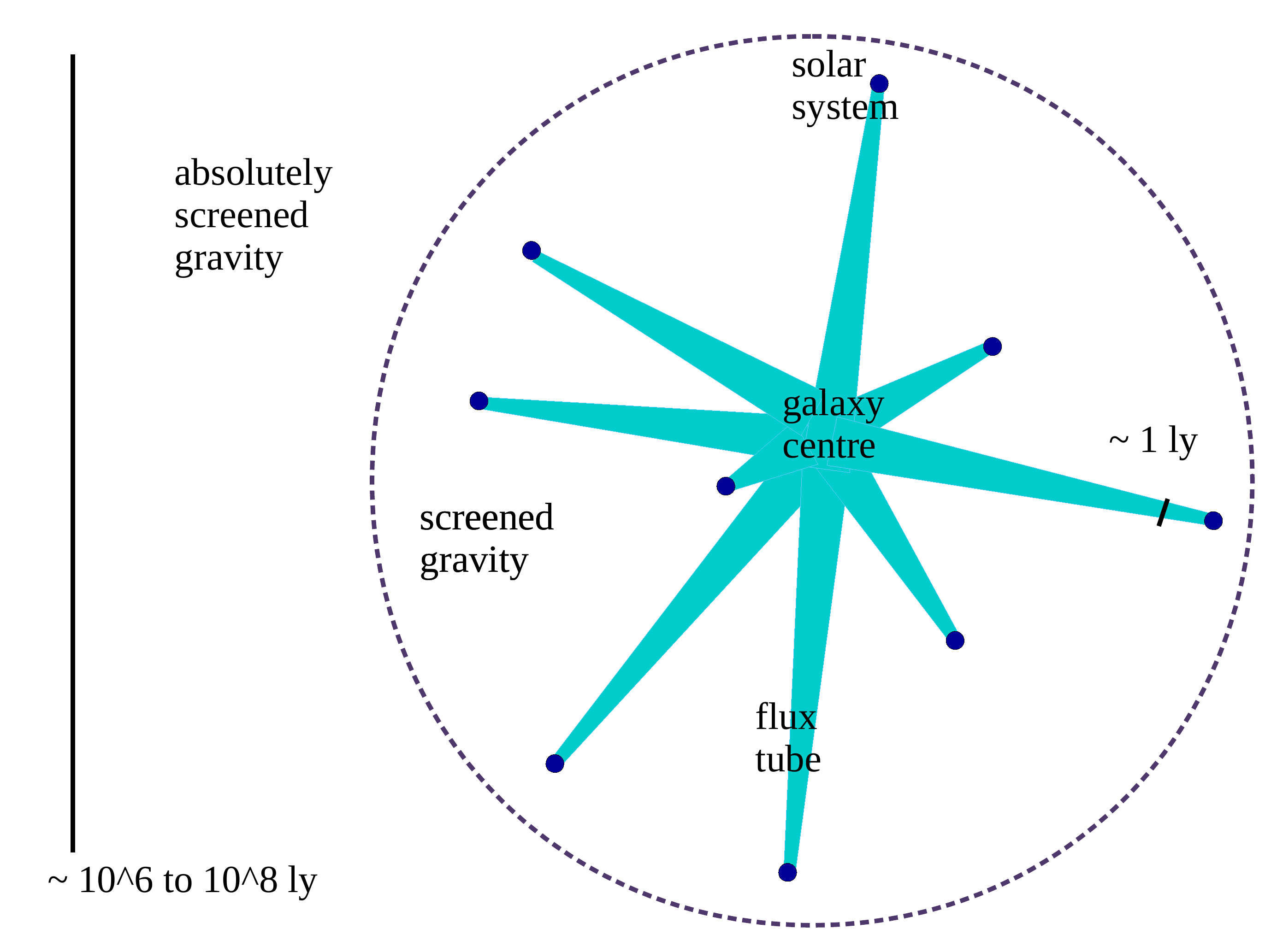}
\caption{(Coulour online) Artist's view (not respecting neither the scales nor the number of stars) of two possible phenomenological models for screening of gravitation, only possible prior to the GW150914 observation at LIGO. Up to solar system radii, $\sim$ 1 ly, Einstein's equations for general gravity would be correct and Newtonian gravity would be a good approximation to gravitation in solar systems. At inter-galactic distances, $\le \ 10^6$ ly screening would lead to MOND with flux tubes of width similar to the size of solar systems, and at distances larger than the galaxy super cluster size, $ \ge 10^8$ ly, screening would wipe out gravitation interactions. }
\label{fig:screening}
\end{figure}

At inter-galactic distances, considering the visible stellar masses distributions and velocities only, the naive Newtonian gravity fails. Two different main solutions have been proposed: dark matter (DM) and modified Newtonian gravity (MOND). While DM has been accumulating observational evidences from lensing and cluster collisions, but has not yet been produced in the laboratory, MOND has been trying to comply with the observations 
\cite{Milgrom:1983ca,Milgrom:1983pn,Milgrom:1983zz,Bugg:2014bka}. Notice MOND corresponds to  a potential stronger than the Coulomb potential, apparently in excess of the Gauss law
\cite{Moody:1993ir} of gravity field flux conservation. In order not to violate the Gauss law, MOND could possibly be due to an effective graviton screening. For distances of the inter-galactic order, a screening similar to the one of the confinement of QCD in strong interactions, where the colour fields are squeezed in flux tubes, could enhance the gravitation force, thus leading to a stronger gravitation force, compatible with MOND.   This is sketched in Fig. \ref{fig:screening}. Notice screening here would not be exactly identical to the one in QCD confinement, where the potential is linear, and thus even stronger than the one in MOND.

Moreover at larger distances than the galaxy cluster scale, the universe expansion has two different solutions as well. One is the dark energy (DE), counter-acting the gravitational pull, and the other is screening. This hypothetical screening is different from the one leading to flux tubes inside the galaxy, since it totally saturates the gravitic potential. It is reminiscent of the screening in effective strong nuclear interactions, mediated by mesons, see Table \ref{tab:screening}, where the strong interaction is short range because the interaction range is of the order of the inverse meson mass. Such a screening should break the Gauss law conservation of the gravitational field flux, eventually suppressing it. In particular, due to the recent theoretical advances in massive gravity, this screening is already studied in Ref. \cite{deRham:2010tw}.
The negative pressure causing the acceleration is due to a condensate of the helicity-0 component of the massive graviton, and the background evolution, in the approximation used in Ref. \cite{deRham:2010tw}, is indistinguishable from the $\Lambda$CDM standard model of cosmology.

Using the relation between the screening mass $m$ and the characteristic distance or penetration length $\lambda$ of Eq. (\ref{eq:screeningmass}), in units of $\hbar=c=1$ a penetration length of 1 m corresponds to a screening mass of 0.197 $\mu$eV and a penetration length of 1 light year (ly) corresponds to a screening mass of $2.08 \times 10^{-23} \text{eV}$. We then arrive at the two different relations for the two hypothetical graviton screening masses. From the solar system characteristic scale \cite{Gott:2003pf} of the order of 1 ly,
\be
m_\text{Gauss preserving} \ll 2 \times 10^{-23} \text{eV} \ .
\label{eq:mond}
\ee 
This mass should squeeze the graviton field in flux tubes, larger than the solar system size, and approximately preserve the Gauss law.
On the other hand from the galaxy typical radius \cite{Gott:2003pf,Juric:2005zr}of $10^6$ ly or super cluster typical radius of the order of $10^8$ ly \cite{Gott:2003pf,Zehavi:2004ii,Cole:2005sx,Springel:2006vs}, we arrive at the absolute screening mass of the order of,
\be
m_\text{absolute screening} \simeq 2 \times 10^{-29} \text{eV} \  \text{to}  \ 2 \times 10^{-31} \text{eV} \ .
\label{eq:expansion}
\ee
Thus, already inside the galaxy (if we assume a screening model of MOND) or beyond the super clusters (if we assume a screening model of the universe expansion), see Fig. \ref{fig:screening} the gravitational field would be screened, and the gravitational wave would be unable to propagate.

However, due to the observed gravitational wave GW150914 at LIGO, assuming the general gravity equations and computer simulations at short distances are correct, then the signal is clearly produced by an extragalactic source. The source of the gravitational wave GW150914 \cite{Abbott:2016blz,TheLIGOScientific:2016wfe} has a luminosity distance of 410 $+160 \atop -180$ Mpc = $=1.3 { +0.6 \atop -0.5 } \times 10^9$ ly from Earth,
corresponding to a redshift of 0.09 $+0.03 \atop -0.04$. Using the linear approximation of the redshift to distance relation, we get a distance of  $=1.2 { +0.4 \atop -0.5 } \times 10^9$ ly from Earth, very close to the luminosity distance.
This means the source is not only extragalactic, it is well beyond our local supercluster with radius of $5.5 \times 10^7$ ly only. But it is close enough that the redshift is still approximately linear with distance, which does not depend strongly on the cosmological model. 

Assuming the signal is able to propagate with no attenuation, then general gravity, which preserves Gauss law, is correct at least up to this distance. This imposes an upper bound for a screening graviton mass of,
\be
m \ll  1.6 \times 10^{-32} \text{eV}  \ ,
\label{eq:GW}
\ee
where the screening mass should clearly be smaller that this bound, otherwise some attenuation effect would already be visible at the distance to the GW150914 source. 
It is important to notice that this bound is for the absolute type of screening only, similar to the confinement screening in QCD, where the wave propagation is damped in the vacuum.

The bound for a Yukawa-like screening, as in nuclear and weak interactions, is much looser, since the corresponding graviton waves propagate through the vacuum, and only the dispersion relation is changed \cite{Will:1997bb}. This bound was already determined by the LIGO Scientific and the Virgo Collaborations \cite{TheLIGOScientific:2016src}, who determined,
\be
m \leq 1.2 \times 10^{-22} \text{eV} \  ,
\label{eq:LIGOSciVirgo}
\ee
 at 90 \% confidence.

It seems clear a screening-driven MOND, with screening mass in Eq. (\ref{eq:mond}), is absolutely excluded by the result in Eq.(\ref{eq:GW}). Moreover, a universe expansion due to the screening of the gravitational interaction between superclusters, with screening mass in Eq. (\ref{eq:expansion}) remains possible if the screning is Yukawa-like with the bound in Eq. (\ref{eq:LIGOSciVirgo}).

Notice the visible universe radius is only $\sim 30 \times$ the distance to the GW150914 source. In the same sense, our bound is getting close to the Hubble constant in eV units, $H_0 \sim 1.5 \times 10^{-33}$ eV  \cite{Xu:2013ega}. As noted in the review \cite{deRham:2014zqa} a graviton mass smaller to this bound would be unobservable in the visible Universe and such a mass would thus loose its phenomenological interest.

Let us now compare our bound with other bounds for the graviton mass or for the mass scale of gravity screening.
From the observation of gravitational waves as we did here, two different sorts of bounds have been predicted. Comparing the speed of light with the speed of gravitational waves from a same source, bounds of $m <   10^{-23}$ eV would be expected form supernovae observations \cite{Will:2014kxa}. With similar ideas as in this work, from the observation of spiralling massive objects, it was expected  that bounds of $m <   10^{-29}$ eV could be extracted in advanced LIGO\cite{Will:2014kxa} . Notice our bound turns out to be tighter by a factor of 1000 than the anticipated bound in Ref. \cite{Will:2014kxa}. Our bound in Eq. (\ref{eq:GW}) is close to the one combining the Lunar Laser Ranging (LLR) experiment \cite{Williams:2004qba} with the present massive gravity theories \cite{deRham:2014zqa}.
It is interesting that a numerical work with black holes already placed a bound on the graviton mass. With the numerical study of both Schwarzschild and slowly rotating Kerr black holes \cite{Brito:2013wya}, it was shown they are unstable for graviton masses $m < 5 \times 10^{-23}$ eV.

To conclude, the event GW150914 observed at LIGO with 5.1 $\sigma$ significance, deeply extends our knowledge of gravitation \cite{Calabrese:2016bnu,Evans:2016mta,Berti:2016iki}. GW150914 just tested general gravity from the size $\sim 10^5 m $ of the black hole merger \cite{Pretorius:2005gq,Campanelli:2005dd,Baker:2005vv,Berti:2007fi,Sperhake:2008ga} to the distance from the black hole merger to the Earth, of  $ \sim 1.3  \times 10^9$ ly. 
Unlike microscopic QCD, nuclear forces and the weak sector of the Standard Model, gravitational wave observations may provide evidence that gravitation has no screening mechanism up to larger than the typical scale of superclusters. In what concerns damping as in QCD-like screening the bound is already very tight, as in Eq. (\ref{eq:GW}). In what concerns the modification of the dispersion relation due to Yukawa-like screening, the present bound in Eq. (\ref{eq:LIGOSciVirgo}) could further be tightened with more gravitational wave observations. 
This provides a tighter bound on screening, and further constrains alternative models to dark matter and dark energy. We are even more confident in Einstein's general gravity \cite{Einstein:1915}, although it still remains to be quantized.

Nevertheless, it would be interesting to extend massive gravity simulations of black holes \cite{Brito:2013wya} to compute numerically black hole mergers, and the resulting gravitational wave. Notice the GW150914 signal, including the luminosity distance, was fitted assuming general relativity, with no screening. A hypothetical screening could damp the luminosity,  and then the effective distance to the GW150914 signal could be shorter than the one reported. This would change our conclusions and relax our bound.

\vspace{10pt}
{\em Acknowledgements -}
P. B. is very grateful to Claudia de Rham for discussions on massive gravity and to David Bugg for discussions on MOND; is thankful to Ana Mourão, George Rupp, José Lemos and Vítor Cardoso for discussions on black holes, gravitation and cosmology; and acknowledges the support of CFTP (grant FCT UID/FIS/00777/2013).


\end{document}